\documentclass{sig-alternate}
\usepackage{pifont}
\usepackage{paralist}
\usepackage{hyperref}
\usepackage{times}

\def\isabridged{1}  
\let\isabridged\undefined 

\newif\ifabridged
\newif\ifnotabridged

\ifdefined\isabridged
\abridgedtrue
\fi

\ifabridged\notabridgedfalse
\else\notabridgedtrue
\fi

\newcommand{\tool}{\textsc{C-FLAT}}
\newcommand{\dOne}{\ding{182}}
\newcommand{\dTwo}{\ding{183}}
\newcommand{\dThree}{\ding{184}}
\newcommand{\dFour}{\ding{185}}
\newcommand{\dFive}{\ding{186}}
\newcommand{\dSix}{\ding{187}}
\newcommand{\dSeven}{\ding{188}}

\newcommand{\adv}{\ensuremath {\mathcal Adv}}
\newcommand{\prv}{\ensuremath {\mathcal Prv}}
\newcommand{\vrf}{\ensuremath {\mathcal Ver}}
\newcommand{\auth}{\ensuremath {\mathcal Auth}}
\newcommand{\func}{\ensuremath {\mathcal H}}

\newcommand{\materials}{\url{https://goo.gl/pTiVdU}}

\def\sharedaffiliation{%
\end{tabular}
\begin{tabular}{c}}
\makeatletter
\def\titlenote{\@ifnextchar[\@xtitlenote{\stepcounter\@mpfn
\global\advance\titlenotecount by 1
\ifnum\titlenotecount=1
      {$^\ast$}
\fi
\ifnum\titlenotecount=2
      {$^\dagger$}
\fi
\ifnum\titlenotecount=3
      {$^\ddagger$}
\fi
\ifnum\titlenotecount=4
      {$^\S$}
\fi
\ifnum\titlenotecount=5
      {$^\P$}
\fi
\@titlenotetext
}}
\makeatother

\ifnotabridged
\makeatletter
\def\@copyrightspace{\relax}
\makeatother
\fi

\begin{document}

\CopyrightYear{2016} \setcopyright{acmcopyright}
 \conferenceinfo{CCS '16,}{October 24--28, 2016, Vienna, Austria.}
 \isbn{978-1-4503-4139-4/16/10}\acmPrice{\$15.00}
 \doi{http://dx.doi.org/10.1145/2976749.2978358} 

\clubpenalty=10000 
\widowpenalty = 10000

\title{C-FLAT: Control-Flow Attestation \\ for Embedded Systems Software}

\numberofauthors{1}
\author{
\alignauthor
Tigist Abera\textsuperscript{$^1$,\titlenote{Author names are listed in alphabetical order.}},
N. Asokan\textsuperscript{$^2$}, Lucas Davi\textsuperscript{$^1$},
Jan-Erik Ekberg\textsuperscript{$^3$},\\ Thomas Nyman\textsuperscript{$^{2,3}$}, Andrew Paverd\textsuperscript{$^2$}, 
Ahmad-Reza Sadeghi\textsuperscript{$^1$},  Gene Tsudik\textsuperscript{$^4$}\\
\sharedaffiliation
      \affaddr{\textsuperscript{$^1$}Technische Universit\"at Darmstadt, Germany}\\
			\affaddr{\{tigist.abera,lucas.davi,ahmad.sadeghi\}@cased.de}\\
\sharedaffiliation
      \affaddr{\textsuperscript{$^2$}Aalto University, Finland}\\
			\affaddr{\{n.asokan,thomas.nyman,andrew.paverd\}@aalto.fi}\\
\sharedaffiliation
      \affaddr{\textsuperscript{$^3$}Trustonic, Finland}\\
			\affaddr{\{jan-erik.ekberg,thomas.nyman\}@trustonic.com}\\
\sharedaffiliation
      \affaddr{\textsuperscript{$^4$}University of California, Irvine, USA}\\
			\affaddr{gts@ics.uci.edu}\\
}

\maketitle
\begin{abstract}
Remote attestation is a crucial security service particularly relevant to increasingly 
popular IoT (and other embedded) devices. It allows a trusted party (verifier) to learn 
the state of a remote, and potentially malware-infected, device (prover). Most existing approaches 
are static in nature and only check whether benign software is {\em initially} loaded 
on the prover. However, they are vulnerable to runtime attacks that hijack the application's 
control or data flow, e.g., via return-oriented programming or data-oriented exploits.

As a concrete step towards more comprehensive runtime remote attestation, 
we present the design and implementation of {\bf C}ontrol-{\bf FL}ow {\bf AT}testation (\tool) that 
enables remote attestation of an application's control-flow path, without requiring the source code. 
We describe a full prototype implementation of \tool\ on 
Raspberry Pi using its ARM TrustZone hardware security extensions.
We evaluate \tool's performance using a real-world embedded (cyber-physical) 
application, and demonstrate its efficacy against control-flow hijacking attacks.
\end{abstract}

\keywords{remote attestation; control-flow attacks; embedded system security}
\section{Introduction} \label{sec:intro}
\noindent
Embedded systems are becoming increasingly ubiquitous, permeating many aspects of  
daily life. They are deployed in automotive, medical, industrial, household and office settings, 
smart cities and factories, as well as in critical infrastructures. 
Unfortunately, this increased popularity also leads to numerous security 
issues~\cite{viega-sp-2012}. Securing embedded devices is challenging, particularly because they are 
special-purpose and resource-constrained~\cite{Kocher:sec-embedded}. 

{\em Remote attestation} is a means of verifying integrity of software running on a remote device. 
Typically, it is realized as a challenge-response protocol allowing a trusted verifier to obtain an 
authentic, and timely report about the software state of a (potentially untrusted and infected) remote device -- a prover. 
Based on that report, a verifier checks whether prover's current state is trustworthy, i.e.,
whether only known and benign software is loaded on the prover. 

The standard trust assumption needed for authenticity of the attestation report 
requires the existence of some trusted component -- called a \emph{trust anchor} -- on the prover. 
Moreover, most attestation schemes assume malicious software (i.e., remote malware infestations) 
as their adversarial model. Prominent examples of trust anchors are secure components 
such as a Trusted Platform Module (TPM). 
Although available on many laptops and desktops, TPMs
are too complex and expensive for deployment on low-end embedded devices. 
Ideally, trust anchors for small embedded devices should be light-weight, i.e., 
require minimal hardware features and assumptions, in light of recent proposals 
such as SMART~\cite{smart}, SANCUS~\cite{sancus}, and Intel's TrustLite~\cite{trustlite}. 
The next generation of ARM Microcontrollers (MCUs)~\cite{ARMv8-M}  
will feature TrustZone-M, a lightweight trust anchor.

Most current remote attestation approaches are {\em static} in nature. 
In such schemes, the prover's report is typically authenticated by means of a cryptographic signature or a 
MAC computed over the verifier's challenge and a \emph{measurement} (typically, a hash) of the binary 
code to be attested. However, static attestation, though efficient, only ensures integrity of binaries
and {\bf not} of their execution. In particular, it does not capture software attacks that hijack 
the program's control flow~\cite{eternal-war}. These attacks tamper with the state information on the application's 
stack or heap to arbitrarily divert execution flow. State-of-the-art memory corruption attacks 
take advantage of code-reuse techniques, such as return-oriented programming, 
that dynamically generate malicious programs based on code snippets (gadgets) of benign code 
\emph{without} injecting any malicious instructions~\cite{rop-journal}.
As a result, the measurements (hashes) 
computed over the binaries remain unchanged and the attestation protocol succeeds, even though the prover is no 
longer trustworthy. These sophisticated exploitation 
techniques have been shown effective on many processor architectures, such as Intel x86~\cite{Sh2007}, 
SPARC~\cite{BuRoShSa2008}, ARM~\cite{Ko2009}, and Atmel AVR~\cite{CaFr2008}.  

The problem arises because static attestation methods do not capture a program's runtime behavior (i.e., 
timely trustworthiness) of the underlying code. In particular, recent large-scale studies have 
shown~\cite{Costin14,Chen16} that embedded software suffers from a 
variety of vulnerabilities, including memory errors (such as buffer overflows), that allow  
runtime exploits.  To be truly effective, an attestation technique should report the prover's 
dynamic state (i.e., its current executions details) to the verifier. As we elaborate in 
Section~\ref{sec:related}, there have been some proposals to enhance and extend static binary 
attestation~\cite{propatt,Franz2004}. However, they either require involvement of an
external trusted third party, or only attest higher level of policies at the Java bytecode layer by 
instrumenting the Java virtual machine. Hence, they do not capture control-flow related attacks 
at the binary level of embedded systems software.

Mitigation of runtime exploitation techniques has been a subject of intensive research.  Prominent 
defenses against control-flow hijacking include: control-flow integrity (CFI)~\cite{Abadi2009}, 
fine-grained code randomization~\cite{Co1993,sok-aslr}, and code-pointer integrity (CPI)~\cite{CPI}. 
However, na\"ively integrating these approaches into remote attestation protocols would provide limited 
state information to the verifier. In particular, these techniques only report whether a control-flow 
attack occurred, and provide no information about the actually executed control-flow path.
Therefore, the verifier can not determine which (of the many possible) paths the prover 
executed. This limitation allows an attacker to undermine such defenses by means of 
so-called data-oriented exploits~\cite{non-control-data}. These attacks corrupt data 
variables to execute a valid, yet unauthorized, control-flow path. A prominent example 
of this attack is the corruption of an authentication variable, allowing the attacker to 
execute a privileged control-flow path. Recently, Hu et al.~\cite{dop} 
demonstrated that such attacks allow Turing-complete malicious computation.


\medskip
\noindent
\textbf{Goals and Contributions.}
This paper proposes \textbf{C}ontrol-\textbf{FL}ow \textbf{AT}testation (\tool), a technique for precise attestation of the 
execution path of an application running on an embedded device. \tool\ complements static attestation 
by measuring the program's execution path at binary level, capturing its runtime behavior. As a new approach, 
\tool\ represents an important and substantial 
advance towards tackling the open problem of runtime attestation. 

\tool\ allows the prover to efficiently compute an aggregated authenticator of the program's 
control flow, i.e., of the exact sequence of executed instructions, including branches and function returns.  
This authenticator represents a  
fingerprint of the control-flow path. It 
allows the verifier to trace the exact execution path in order to determine whether application's 
control flow has been compromised. 
Combined with static attestation, \tool\ can precisely attest embedded software execution so as
to allow detection of runtime attacks. 

In designing \tool, we focus on embedded systems. As the initial proof-of-concept, we attest 
single-thread programs executed by small IoT MCUs, since \tool\ is not meant for arbitrary
complex applications. We discuss the path towards control-flow attestation of MCUs in 
Section~\ref{sec:towards-mcus}.

The main contributions of this paper are:
\begin{compactitem}
	\item A novel and practical scheme for attestation of the
	application's execution path. In contrast to more
	traditional static attestation, \tool\ captures application's runtime behavior (Section~\ref{sec:design}).
	\item Addressing several challenges unique to control-flow path attestation, such as handling
	loops and call-return matching (Section~\ref{sec:challenges}).
	\item A proof-of-concept implementation that features (i)~a static analysis tool to determine 
	valid control flows of an application, (ii)~a static binary instrumentation tool to extend ARM binaries 
	with \tool\ functionality, and (iii)~a background service implemented as a trusted application, using ARM TrustZone extensions,
	which monitors runtime control flow and generates the attestation response (Section~\ref{sec:implementation}).
	\item A systematic evaluation of \tool\ in the context of \emph{Open Syringe Pump}, a real embedded 
	control application for a widely-used class of \emph{cyber-physical} systems (Section~\ref{sec:evaluation}). 
	We also demonstrate \tool's resilience against various runtime attacks (Section~\ref{sec:security}).
\end{compactitem}
We chose to initially instantiate \tool\ using ARM TrustZone since ARM-based platforms (particularly 
Raspberry Pi) are widely available and popular for building embedded applications, 
e.g., the syringe pump described in Section~\ref{sec:case-study}. Although our 
implementation uses TrustZone-A hardware security extensions available in current 
general-purpose ARM processors, its more lightweight successor TrustZone-M will be available
on ARMv8 MCUs~\cite{ARMv8-M} which are expected to come on the market soon.
\ifnotabridged
This will allow \tool\ to be easily realizable on commercial MCUs.
We elaborate on the use of embedded architectures with small trust anchors, and discuss
further performance improvements in Section~\ref{sec:extensions}.
\else
This will allow \tool\ to be easily realizable on commercial MCUs (see Section~\ref{sec:extensions}).
\fi

\medskip
\noindent
\ifabridged
\textbf{Code Availability and Technical Report.}
\else
\textbf{Code Availability.}
\fi
To enable reproducibility of our results, and to encourage further research in this area, 
the source code for \tool, our use-case cyber-physical programs, and the proof-of-concept 
exploits are available at \url{https://goo.gl/pTiVdU}. 
\ifabridged
We also provide an extended version of this paper as a technical report~\cite{TR-CFLAT}. 
\fi

\section{Problem Setting} \label{sec:background} \label{sec:bg-attacks}
\noindent
Runtime attacks exploit program vulnerabilities to cause malicious and unauthorized program 
actions. The most prominent example is a buffer overflow, allowing the attacker to corrupt 
memory cells adjacent to a buffer. The main target of these attacks is manipulation 
of control-flow information stored on the program's stack and heap. 

\begin{figure}[htbp]
	\centering
		\includegraphics[width=0.75\linewidth]{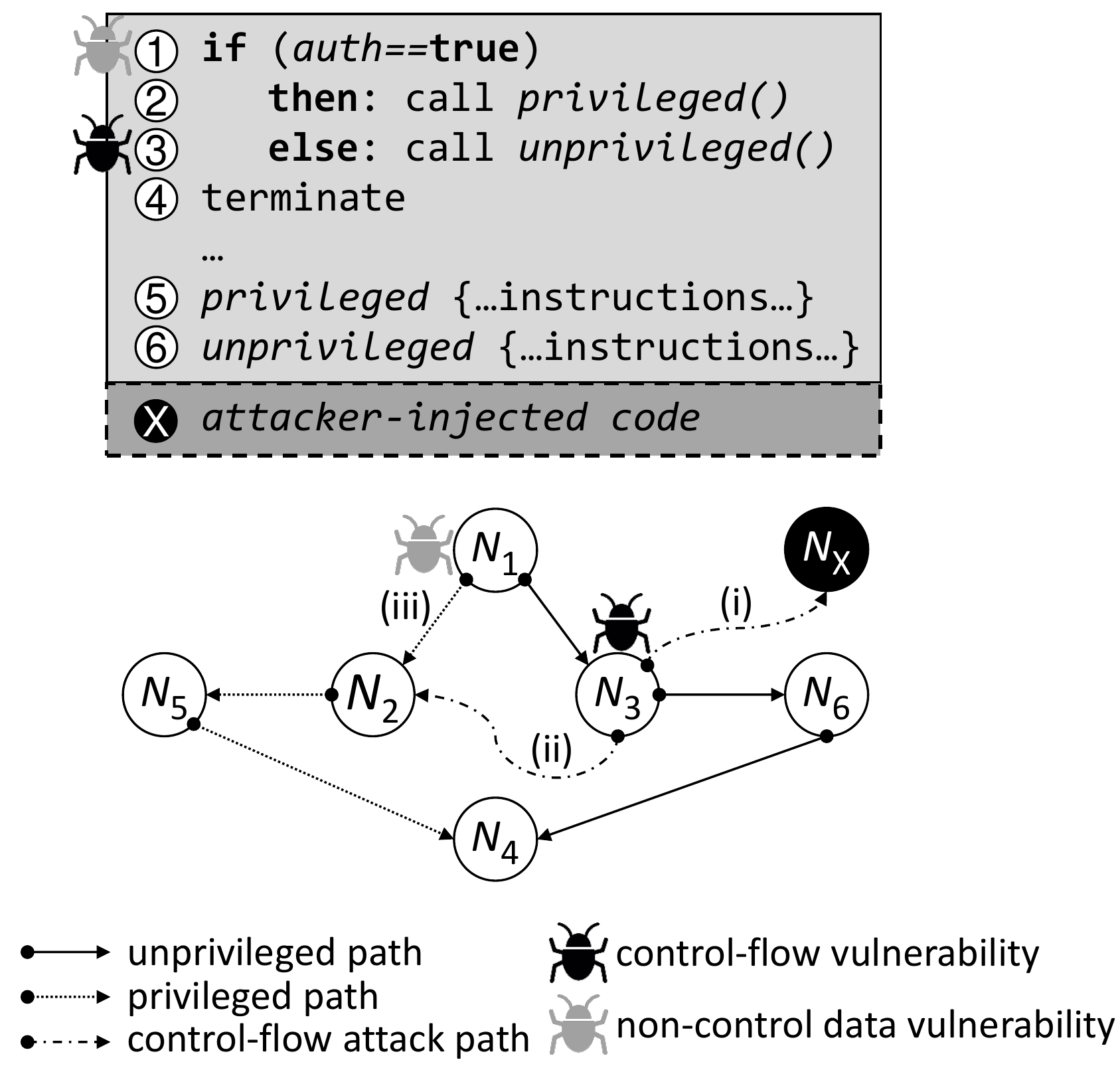}
	\caption{Abstract view of runtime attacks}
	\label{fig:runtime-attacks}
\end{figure}

Figure~\ref{fig:runtime-attacks} shows a generic runtime attack example of a program that
calls either a privileged or non-privileged subroutine based on the 
authentication variable \textit{auth}.\footnote{In general, any program can be abstracted through its corresponding 
control-flow graph (CFG), where nodes represent code blocks and edges control-flow 
transitions.} Line numbers in the pseudo-code map to CFG nodes. 
The example program suffers from a control-flow vulnerability at node $N_3$ allowing the 
attacker to overwrite code pointers that store control-flow information. (Typically, these vulnerabilities allow
the attacker to read from, and write to, the application's memory.) 
Upon encountering the corrupted code pointer, the program's control flow is deviated to either (i)~previously 
injected code (node~$N_X$)~\cite{AlephOne96} or (ii)~existing code (node~$N_2$) such as system 
functions~\cite{solar-ret2libc} or 
unintended code sequences~\cite{Sh2007}. The latter is commonly referred to as 
\emph{code-reuse attack}, one type of which -- called \emph{return-oriented programming} -- allows the attacker to 
generate arbitrary malicious 
program actions based on chaining short sequences of benign code. Since code-reuse attacks 
do not require injection of malicious code, they undermine the widely-deployed security model 
of data execution prevention (DEP)~\cite{DEP} which aims at preventing code injection attacks 
by either marking memory as writable or executable. 


Code-reuse attacks have emerged as the state-of-the-art exploitation technique on various 
processor architectures. Both control-flow integrity (CFI)~\cite{Abadi2009} and code-pointer integrity 
(CPI)~\cite{CPI} aim at mitigating these attacks. While CFI enforces the program always 
following a legitimate path, CPI ensures integrity of code pointers. However, these schemes do 
not cover so-called \emph{non-control-data attacks}~\cite{non-control-data}.  
These attacks corrupt data variables which are 
used to drive the control flow of the program. In the example of
Figure~\ref{fig:runtime-attacks}, node $N_1$ transitions the control flow to either $N_2$ or $N_3$,
based on \textit{auth}. Thus, if the program suffers from a non-control-data vulnerability, 
the attacker can change \textit{auth} from false to true, so that execution 
continues in the privileged path although the user has not been authenticated to execute that path, i.e., 
attack path (iii) in Figure~\ref{fig:overview}.

For the sake of completeness, attacks discussed thus far lead to unintended, yet valid, program flows. However, 
it is also possible to mount pure data-oriented programming (DOP) attacks which only corrupt memory load and 
store operations, without inducing any unintended program flows~\cite{dop}. Consider a program 
that reads from a buffer via a data pointer and sends retrieved data over network. A pure 
DOP attack would only manipulate the data pointer, e.g., to reference a private key. Hence, the 
original program control flow would lead to leakage of the private key without incurring any 
suspicious control flows. Since we focus on control-flow attestation, such pure data-flow attacks 
are beyond the scope of this paper. As shown in Figure~\ref{fig:runtime-attacks}, we focus 
on control-flow related attacks launched either 
by manipulating control-flow information, or non-control-data, 
such as data variables, i.e., the attacks (i)-(iii). 

%
%

\section{System Model} \label{sec:model}
\begin{figure}[htbp]
	\centering
		\includegraphics[width=0.44\textwidth]{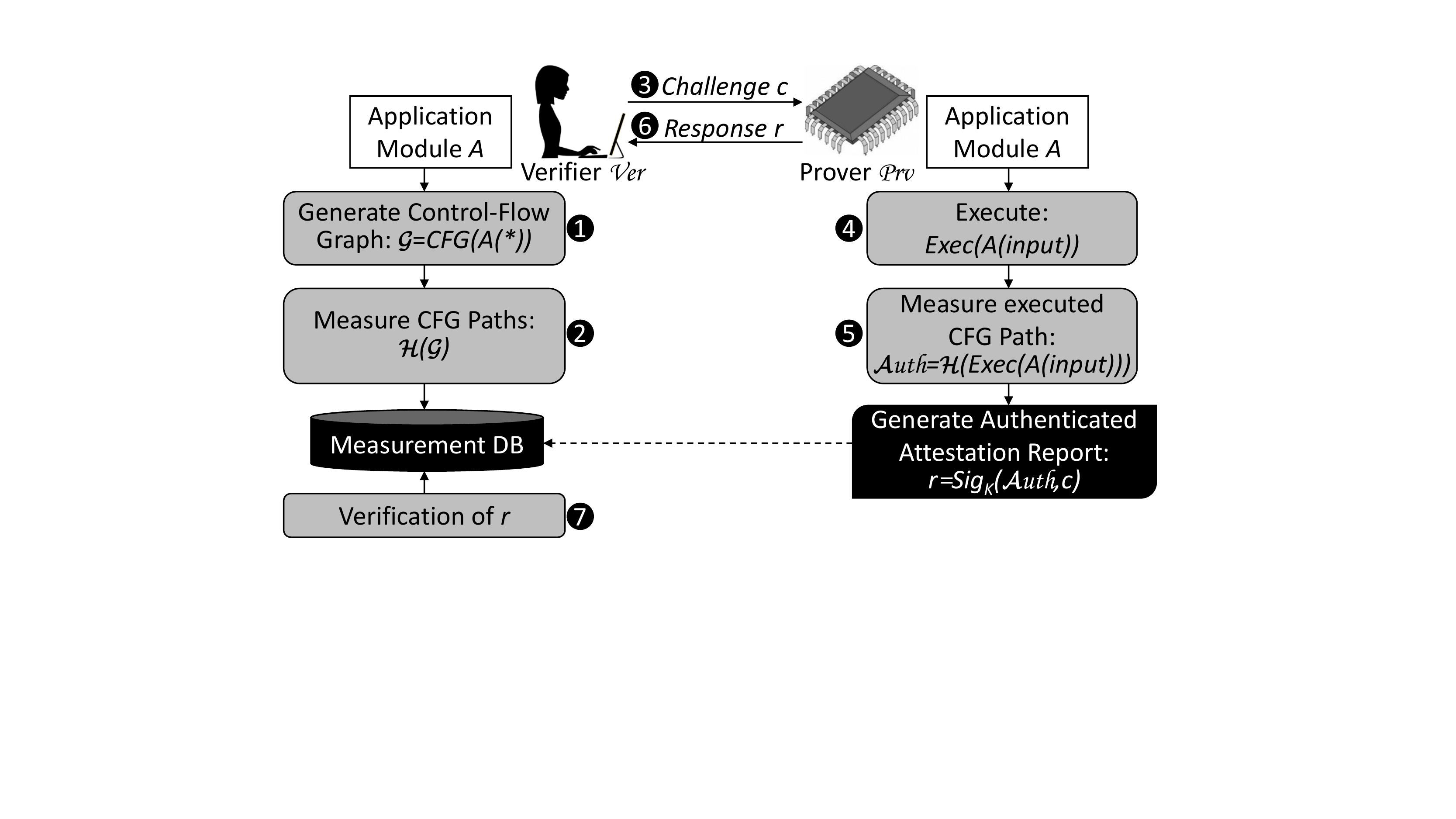}
	\caption{Overview of \tool}
	\label{fig:overview}
\end{figure}

\noindent
Figure~\ref{fig:overview} shows our system model:
the verifier \vrf\ wants to attest runtime control flows of an application module on a 
remote embedded system -- the prover \prv. The application module is typically an 
entire program or a subset thereof, such as a specific function. 
Both \vrf\ and \prv\ are assumed to have access to the binaries of the underlying application.   
%

\subsection{Overview}
\noindent
\tool\ requires \vrf\ to perform offline pre-processing: (1)~generate
the control-flow graph (CFG) of the application module via static analysis (\dOne), 
and (2)~measure each possible control-flow path using a measurement 
function \func, and store the result in a 
measurement database (\dTwo). This needs to be done only once per application module. 
Since \prv\ is assumed to be a low-end anemic embedded device with very limited memory, 
it cannot generate and store the entire CFG. However, this can be easily done by \vrf\ which
has no such resource limitations. 

\noindent {\em NOTE:}
We realize that efficient computation of a generic program's 
CFG and exploration of all possible execution paths is an open problem. However, 
in this paper, we focus on static embedded application software which is typically 
much simpler than programs for general-purpose computers, e.g., the syringe pump 
program we analyze for our proof-of-concept consists of only $13$k instructions.

In \dThree, \vrf\ asks \prv\ to execute the application by transmitting 
a challenge that includes the name of the application module\footnote{In practice, 
the verifier could transmit the actual
application code.} and a nonce to ensure freshness. 
Next, \prv\ initiates execution of the application module (\dFour), while
a dedicated and trusted Measurement Engine computes a cumulative authenticator 
\auth\ of the control-flow path (\dFive). At the end, \prv\ generates the \emph{attestation report} $r=\mbox{Sig}_K(\auth,c)$, 
computed as a digital signature over the challenge $c$ and \auth\ using a key $K$ known only to the 
Measurement Engine. Finally, \prv\ sends $r$ to \vrf\ (\dSix) for validation (\dSeven). 

Since \prv\ attests the application code (with static attestation) and its control flow 
(with \tool), \vrf\ can detect runtime attacks, as discussed in Section~\ref{sec:bg-attacks}. 
Any deviation from the program's legitimate control flow  
results in an unrecognized measurement. Further, non-control data attacks output 
an unexpected, though valid, measurement allowing \vrf\ to detect attacks within the 
application's valid CFG. Static attestation (not shown in Figure~\ref{fig:overview}), 
assures \vrf\ that \prv\ is running the intended application.

\subsection{Requirements and Adversarial Model} \label{sec:adv-model}
\noindent
As expected of an attestation scheme, we require that the attestation report ($r$) must be 
a \emph{fresh} and \emph{authentic} representation of the application's runtime control flows.
The scheme itself must therefore be resilient against replay and masquerading attacks.

We make the following assumptions about the behavior of the adversary, denoted by \adv.

\adv\ can introduce
	arbitrary malware into the prover. However, we rule out physical attacks, which 
	is a standard assumption in all single-prover attestation protocols.  
	We focus on attacks that hijack the execution path of the code to be 
	attested (see Section~\ref{sec:bg-attacks}). 
	Such attacks are based on providing malicious inputs to \prv's public interfaces.
Furthermore, we assume the following \prv\ features:
\begin{compactitem}
	\item Data execution prevention (DEP) to prevent an attacker from injecting and 
	executing malicious code into running processes. All modern platforms -- including embedded 
	systems -- provide hardware to support enforcement of this principle.
	\item A trust anchor that: (1)~provides an isolated measurement engine, which 
	cannot be disabled or modified by non-physical means (i.e., it allows measurements 
	for both the static and control-flow path attestation), and (2)~generates a fresh,
	authenticated attestation report. In Section~\ref{sec:towards-mcus}, 
	we discuss several concrete instantiations of this trust anchor.  
\end{compactitem}
%
These assumptions are in line with all previous work on remote attestation\footnote{Assuming a trust anchor is typical for \emph{remote} attestation protocols.}.
However, we consider a stronger \adv, capable of control-flow hijacking attacks.

\section{C-FLAT Design} \label{sec:design}
%
\noindent In order to perform control-flow attestation, \vrf\ asks 
for a measurement of \prv's execution path. This measurement should  
allow \vrf\ to efficiently determine and verify \prv's control-flow path. It is
clearly infeasible to record and transmit every executed instruction, since that 
would: (1) result in a very long attestation response which \prv\ would have to store,
and (2) require \vrf\ to walk through every single instruction. The same applies to another
intuitive approach that would record and transmit source and target addresses of every 
executed branch, since such instructions frequently occur during program execution. 
To keep the attestation response short and allow fast verification, we propose a cumulative 
hash-based control-flow attestation scheme that builds a hash chain of executed 
control-flow transitions.

\subsection{C-FLAT: High-Level Description} \label{sec:idea}
\begin{figure}[htbp]
	\centering
		\includegraphics[width=0.8\linewidth]{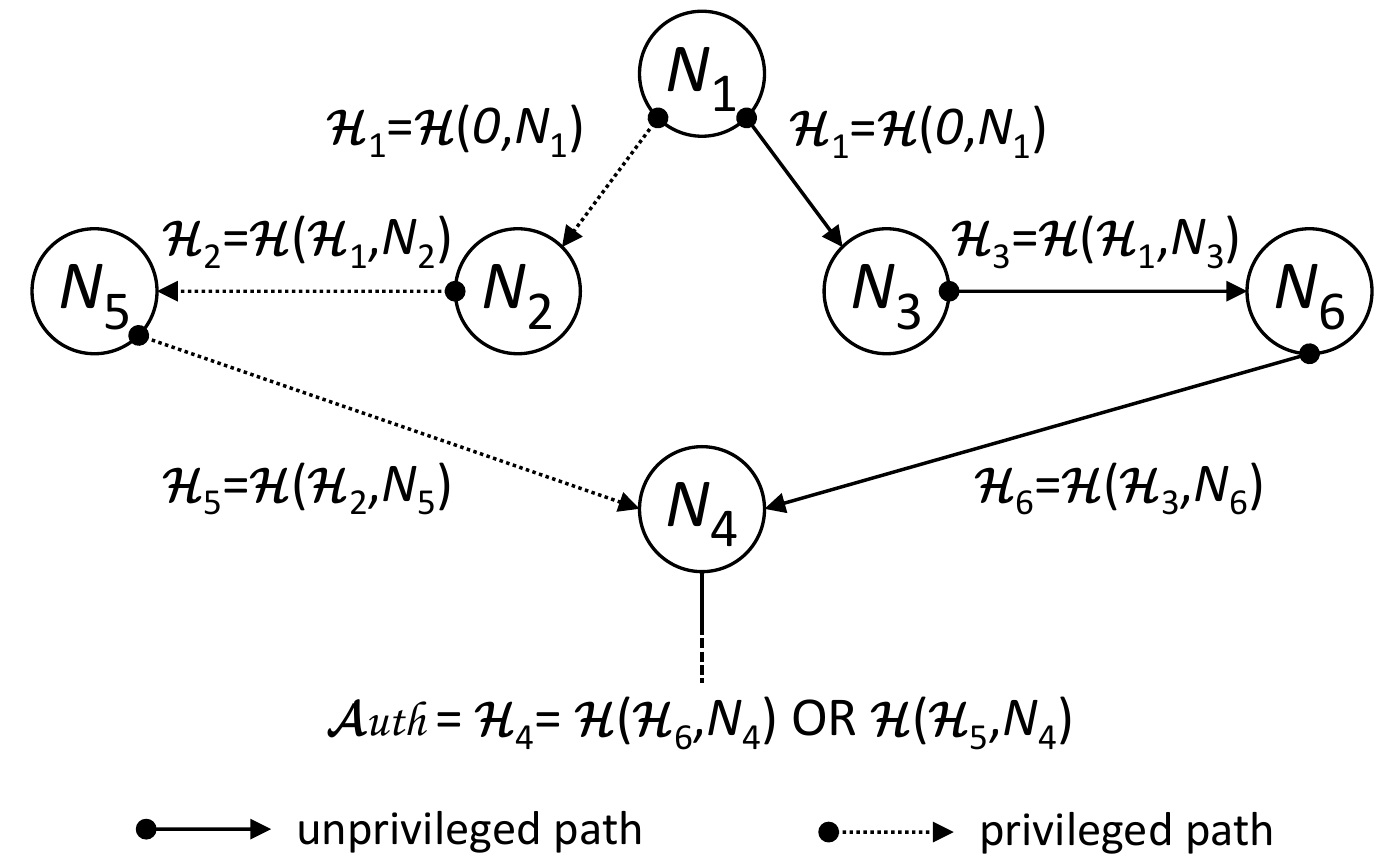}
	\caption{\tool\ control-flow attestation}
	\label{fig:idea}
\end{figure}

\noindent
The main idea is to extend static (hash-based) attestation of 
binary files to dynamic (runtime) control-flow paths. Figure~\ref{fig:idea} illustrates
this idea based on a simple control-flow graph (CFG)  already shown in Section~\ref{sec:bg-attacks}. 
Each CFG node contains a set of assembler instructions, 
and each edge represents a node transition by means of a branch instruction. Depending on 
the desired granularity, the nodes can be (1) entire functions, (2) basic blocks (BBLs) ending in 
an indirect branch, or (3) BBLs ending in any branch instruction, e.g., direct or indirect jump, 
call and return. In this paper, we consider the last case allowing \vrf\ to precisely 
validate the executed control-flow path of an application on \prv's device.

\tool\ depends on every executed branch instruction. It employs a measurement function 
\func\ which takes as input: (1) node ID of the source node $N_i$, and (2) previous 
measurement: $\func_i=\func(\func_{prev}, N_i)$. At the beginning, when no previous 
measurement is available, we start with $\func_{prev}=0$. As a result, 
there is a cumulative measurement for each possible program path, e.g., the privileged path 
outputs $\func_5$, while the unprivileged path leads to $\func_6$ in Figure~\ref{fig:idea}. Any 
unexpected measurement indicates to \vrf\ that an illegal path has been executed. Furthermore, based 
on the reported measurement, \vrf\ can easily determine whether the privileged path has been executed.

Due to its speed and simplicity, we chose the cryptographic hash function 
BLAKE-2\footnote{\url{https://blake2.net}} as \func\ for cumulative hashing that yields \auth.
Hash-based measurements are already deployed in static attestation and 
allow mapping of an input to a unique\footnote{With overwhelming probability.} 
fixed-size result. To generate the final attestation report $r$, \prv\ can use 
a public key signature or a MAC over \vrf's challenge $c$ and \auth. 
(Without loss of generality, we use signatures in the rest of the paper.)
In either case, the secret (private) key is assumed to be protected within the trust anchor.

Obviously, the number of valid measurements depends on the complexity and size of the 
application module, particularly, the number of indirect and conditional branches. 
Indirect branches may target from $1$ to $n$ nodes, and conditional branches target 
$2$ nodes in the CFG. Loops and recursive calls also lead to a high number of 
valid measurements, depending on the number of loop iterations, or recursive 
calls. In Section~\ref{sec:challenges} we address this challenge using an 
approach for efficient loop and recursion handling, allowing us to limit the number of 
possible (legal) measurements.

\subsection{Challenges} \label{sec:challenges}
\noindent
There are several challenges in instantiating \tool. First, na\"ively applying it to arbitrary code 
containing loops can lead to a combinatorial explosion of legal \auth\ values, since each 
execution involving a distinct number of loop iterations would yield a different \auth\ value. 
Furthermore, a static CFG does not capture call-return matching, e.g., a subroutine might
return to various call locations. This would necessitate allowing too many possible \auth\ 
values that could be exploited by \adv~\cite{rop-cfi-davi}. 

\begin{figure}[htbp]
	\centering
		\includegraphics[scale=0.39]{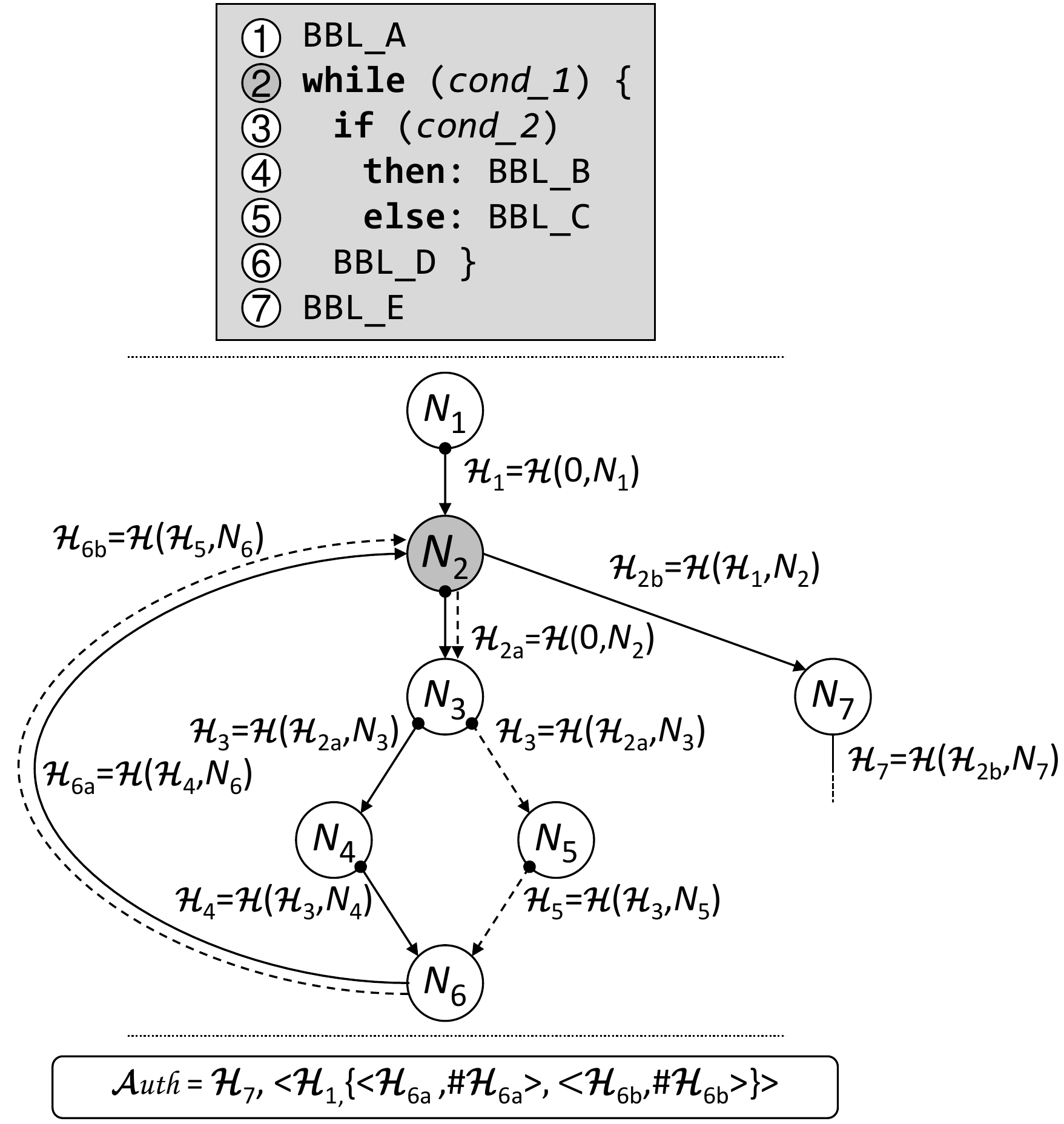}
	\caption{Loop Handling in \tool}
	\label{fig:loops}
\end{figure}

\noindent\textbf{Loops.}
\noindent
Figure~\ref{fig:loops} depicts a CFG and its corresponding pseudo-code for a classic 
\textit{while} loop, which contains an \textit{if-else} statement.
Note that conditional statements (lines~2 and~3) introduce nodes ($N_2$, $N_3$) with 
multiple outgoing edges. Hence, based on a condition check, they follow one of these edges. 
The main challenge is that the cumulative hash computed at $N_2$ is different at each loop iteration, since 
it subsumes the cumulative hash produced after the previous iteration. For an application that contains 
multiple (and even nested) loops, the number of legal \auth\ values grows exponentially,
making control-flow attestation cumbersome. 

Our approach to tackling this problem is based on handling a loop as a sub-program. 
We measure each loop execution separately and merge its cumulative value with 
that of the previous execution, at loop exit. Consider the example in Figure~\ref{fig:loops}: 
first, we identify the loop starting node by means of static analysis -- $N_2$. 
Second, we initiate a new computation when the loop is entered, i.e., 
$\func_{2a}=\func(0,N_2)$. To avoid losing the previous value, we store $\func_1$. 

Our example also has an \textit{if-else} statement within the loop. It diverts control flow at 
$N_3$ to either $N_4$ or $N_5$, depending on \textit{cond\_2}. Consequently, each loop 
iteration can either output $\func_{6a}$ (solid line) or $\func_{6b}$ (dashed line). 
Upon loop exit, 
it is also desirable to attest the number of times a loop is executed. To do so, 
we track each loop measurement separately, by storing the number of times 
a distinct measurement was encountered at $N_2$, and including this result 
$\#\func_{6a},\#\func_{6b}$ in \auth\ 
where $\#\func_{i}$ reflects the number of loop iterations for each possible path. 
This ensures that every loop iteration is attested. However, the size of 
\auth\ is now expanded by the number of different paths inside the loop. 
In Section~\ref{sec:attestation-results}, we demonstrate some actual \auth\ sizes for 
several sample runs of the syringe pump application.

Upon loop exit, we simply take the recorded measurement of the execution right before the loop 
was entered ($\func_1$) for the computation of $\func_{2b}$. The measurement for 
$\func_{2b}$ does not capture whether the loop has executed. However, observed loop path 
measurements are reported in \auth. 
We denote these with 
$<\func_i,\#\func_i>$ to report the loop path measurement and the number of times this measurement 
has been observed. Since a program might enter the loop at different execution stages, 
we need to separate loop counters for each stage. Otherwise, \vrf\ will not be able to determine
the time when the loop has been entered. For this, we maintain 
a reference that indicates the program stage at which loop measurements were taken. This 
is achieved by simply taking the previous measurement right before the loop is entered, e.g., $\func_1$
in Figure~\ref{fig:loops}.

This general approach also supports nested loops and recursive function calls, since each starting
node initiates a separate measurement chain and the result at loop exit (or recursion return) is
reported as separate measurements in 
\auth.
We also perform call-return matching
on \prv\ in order to distinguish multiple instances of a particular loop from each other. This
includes recursive calls to subroutines that contain loops. Such loops occur in the CFG only
once but at runtime multiple loop instances might exist concurrently. In such cases, each distinct
instance is represented separately in \auth.
\ifnotabridged
Call-return matching also allows us to correctly handle subroutine calls that occur within a loop
body. When determining the loop exit node, we identify the basic block from which branches to the
loop entry node is no longer possible during benign execution. In other words the execution of the
exit node, or subsequent basic blocks residing in memory after the exit node in, indicates the end
of a particular instance of the loop. However, consider the scenario described by
Figure~\ref{fig:call-ret-loops}. It depicts a loop that will branch to subroutine \texttt{f} from
$N_3$ on each iteration. Note that the subroutine entry $N_6$ resides in memory after the loop, and
would therefore erroneously be considered to signify the end of the loop. By performing call-return
matching, we can detect when execution leaves the context of the loop (i.e.  the routine in which
the loop resides) and exit nodes need not be tested for until execution of the loop code resumes. 

\begin{figure}[htbp]
	\centering
		\includegraphics[scale=0.39]{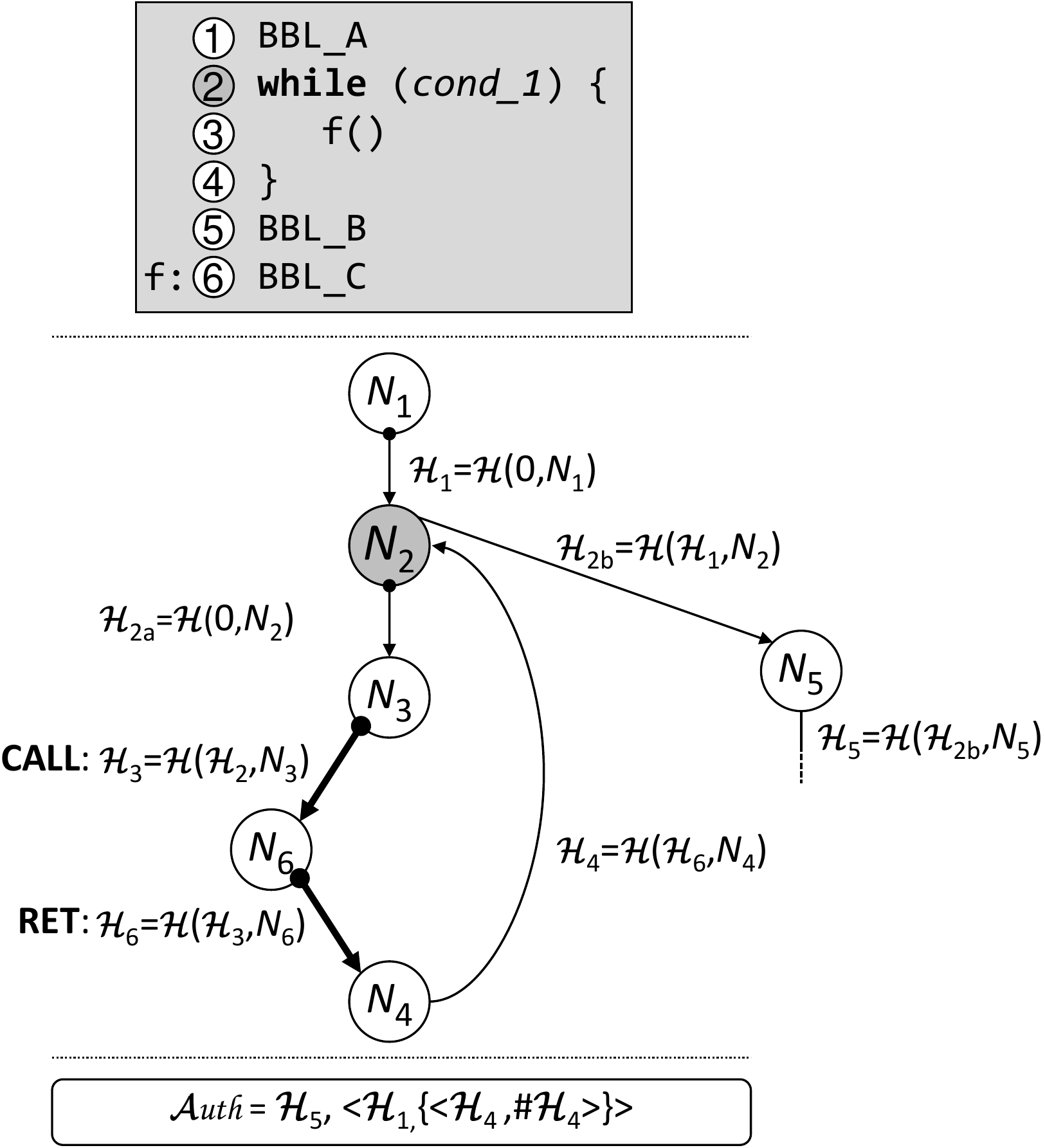}
	\caption{Subroutine calls within a loop body}
	\label{fig:call-ret-loops}
\end{figure}
\else
For further details we refer the reader to the accompanying technical report~\cite{TR-CFLAT}.
\fi

\begin{figure}[htbp]
	\centering
		\includegraphics[scale=0.41]{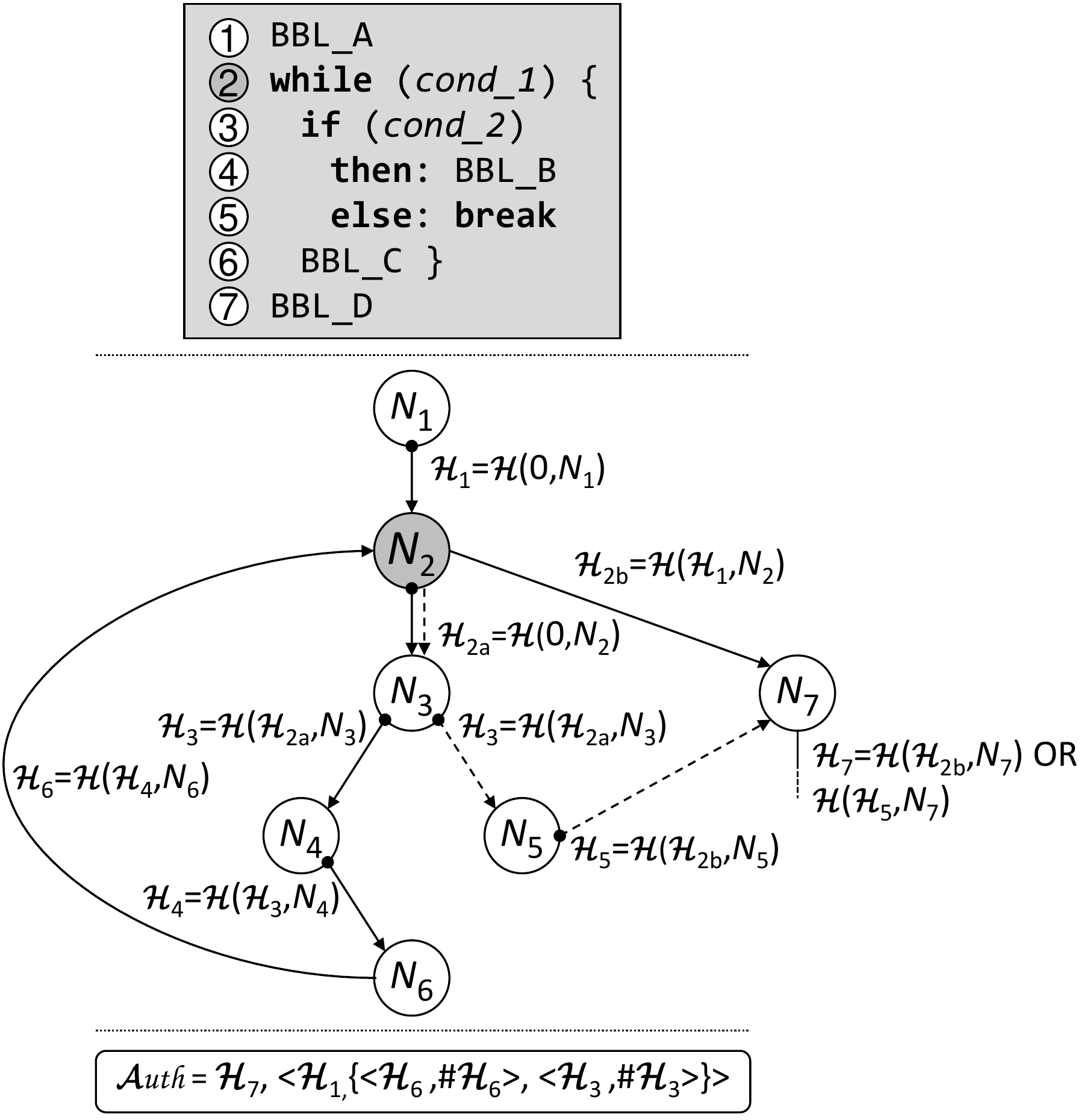}
	\caption{Handling Break Statements in \tool}
	\label{fig:breaks}
\end{figure}

\noindent\textbf{Break Statements.}
Another related challenge arises when \emph{break} statements occur inside loops or switch-case 
statements. Once encountered during program execution, they immediately terminate the innermost loop. 
Figure~\ref{fig:breaks} shows a code example and its associated CFG for such a 
case. Compared to the previous loop example in Figure~\ref{fig:loops}, node $N_5$ 
now executes a \emph{break} instruction, where the path leading to the \emph{break} is indicated 
by a dotted line.
 
We need to treat \emph{break} statements inside loops as special loop exit nodes with a slightly
modified measurement function, for the following reason. Typically, loop exits only occur at the
conditional check at node $N_2$. However, \emph{break} in $N_5$ will not return back to the
conditional statement, but immediately to $N_7$. Hence, we slightly extend the original measurement
function to capture the fact that the measurement right before the loop was $\func_1$ and the loop
has been terminated over the \emph{break} statement. To do so, the measurement at $N_5$ indicates
that the loop has been entered at $N_2$ with the previous hash value $\func_1$ ($\func_{2b}$) and
terminated via $N_5$. The same principle is applied to goto statements that terminate a loop,
e.g., a goto from a nested loop that also terminates an outer loop.

\begin{figure}[htbp]
	\centering
		\includegraphics[scale=0.39]{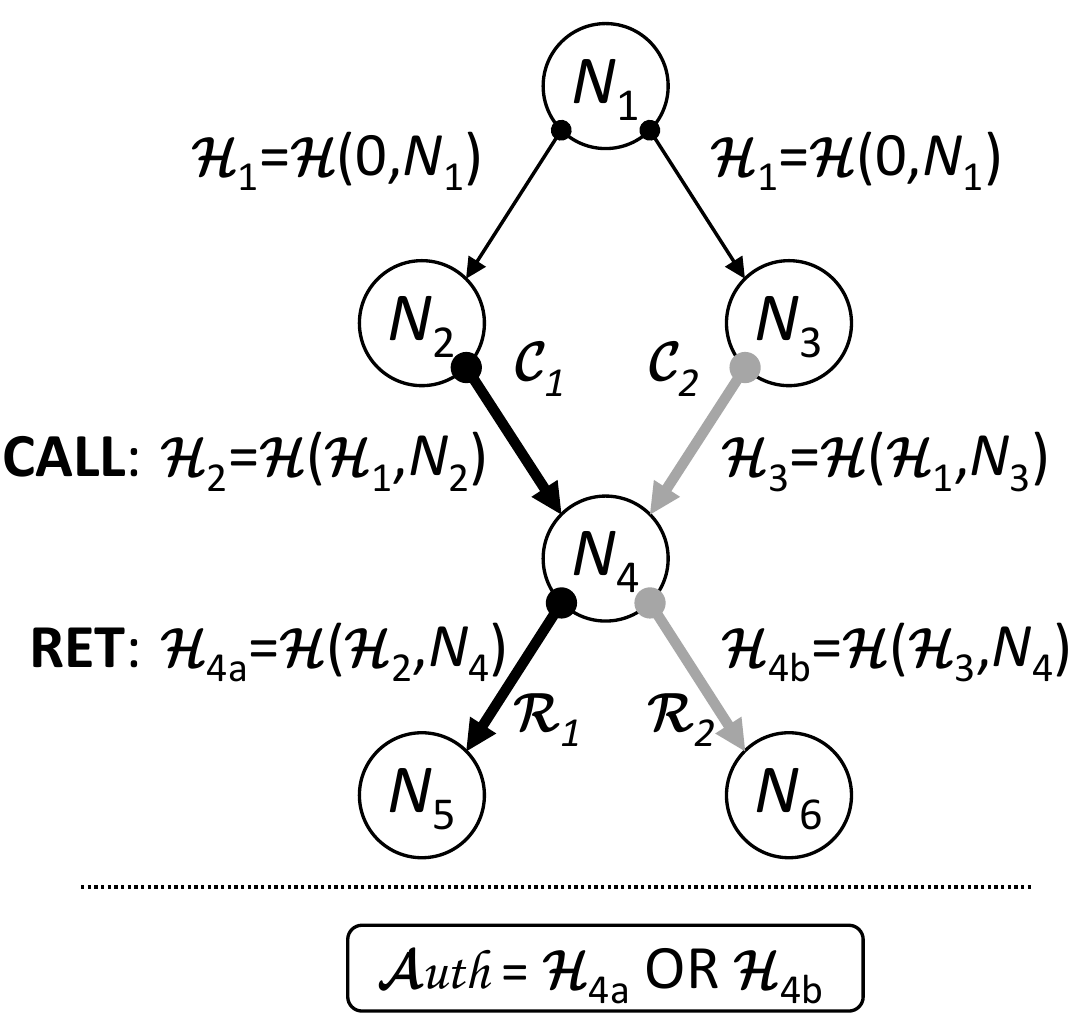}
	\caption{Call-return matching}
	\label{fig:call-ret}
\end{figure}

\noindent\textbf{Call-Return Matching.}
Function calls and returns pose another challenge, especially, when a subroutine can
be invoked from  multiple calling locations. Figure~\ref{fig:call-ret} depicts a sample scenario 
where two function calls ($N_2$, $N_3$) call subroutine $N_4$. In the static CFG, the return 
instruction at $N_4$ can either target $N_5$ or $N_6$. However, if the previous flow is not 
considered, \adv\ can execute a malicious path, e.g., $N_2~\mapsto~N_4~\mapsto~N_6$. 
As recent work demonstrates, such malicious flows allow for a variety of control-flow 
attacks~\cite{rop-cfi-davi}.

To cope with this issue, we index call and return edges during static analysis. In 
the CFG of Figure~\ref{fig:call-ret}, the call from $N_2\mapsto~N_4$ is indexed with 
$C_1$ and the corresponding return edge $N_4~\mapsto~N_5$ -- 
with the same index marked as $R_1$. Hence, the only \auth\ values \vrf\ 
considers legal are $\func_{4a}$ and $\func_{4b}$.

\section{Implementation} \label{sec:implementation}
\noindent
\interfootnotelinepenalty=100000
This section presents our prototype instantiation of \tool\ and discusses its key implementation aspects.

\subsection{Architecture of Proof-of-Concept} \label{sec:poc}
\noindent
To demonstrate the feasibility of \tool, we prototyped it on a popular embedded platform. As
discussed in Section~\ref{sec:adv-model}, remote attestation requires a trust anchor on \prv.
Because lightweight hardware trust anchors for small embedded devices, such as Intel's TrustLite or
ARM's TrustZone-M, are not currently commercially available we opted for a more powerful ARM
application processor that features TrustZone-A~\cite{ARM-TrustZone} security extensions for our
prototype. Specifically, we chose Raspberry Pi 2 as the underlying embedded device platform. In our
experimental setup, Raspberry Pi 2 is set up to execute a \emph{bare metal}, monolithic,
single-purpose program in the ``\emph{Normal World}''~\cite{ARM-TrustZone} of the ARM core,
\emph{without} a separate OS. In Section~\ref{sec:extensions}, we explain why 
the same approach is also applicable in systems where applications run on top of an embedded OS.

\begin{figure}[htbp]
	\centering
		\includegraphics[width=0.9\linewidth]{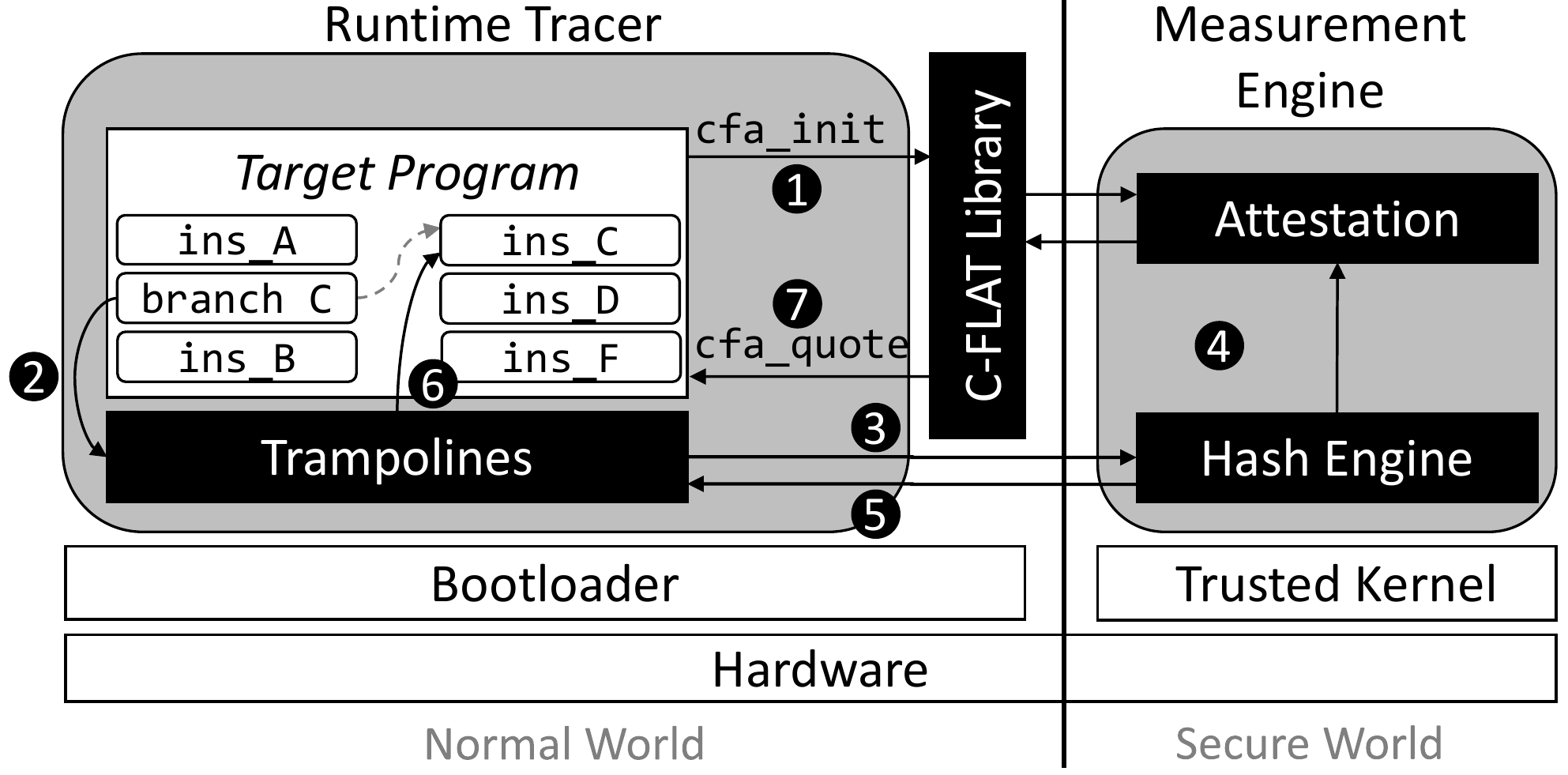}
	\caption{Proof-of-concept architecture}
	\label{fig:architecture}
\end{figure}

The prototype has two main components: i)~a program analyzer to compute legal measurements of the
target program, and ii)~a \emph{Runtime Tracer} and an isolated, trusted \emph{Measurement Engine}
to trace and measure the runtime control-flow path. The former can be
realized either as: (1) a static binary analyzer that generates the program's CFG by identifying
basic blocks and their connection through branch instructions, or (2) a dynamic analyzer that
produces valid measurements for a set of inputs by tracing execution of the target program. We opted
for a dynamic analyzer. We also developed a binary rewriting tool for ARM binaries used to
instrument the target program in order to allow the Runtime Tracer to intercept all branches during
execution. Figure~\ref{fig:architecture} shows the \tool\ prototype architecture. Black boxes depict
\tool\ system components. Below we describe the operation of the prototype.


\noindent\textbf{Instrumentation Phase.}
Initially the program binary is analyzed and instrumented using our instrumentation tool. During
the analysis, information about direct branches and loops is collected from the binary, and stored
in the \emph{branch table} and \emph{loop table} data structures included as part of the Runtime Tracer in the software image loaded
onto Raspberry Pi 2. At runtime, these data structures are made available in read-only memory to the
Runtime Tracer and Measurement Engine.

During instrumentation, each control-flow instruction in the target program is modified to pass
control to \emph{Trampolines} belonging to the Runtime Tracer. Further details about the
instrumentation tool can be found in Section~\ref{sec:instr}. We note that the analysis and
transformations done to the target program could also be performed at program load time, either by a
trusted bootloader, or a program loader in configurations where an embedded operating system is
present on the device.

\noindent\textbf{Runtime Operation.}
The \tool\ Library (\dOne) serves as a mediator between the attested program on \prv\ and the
Measurement Engine. It provides an API that allows software on \prv\ to initiate the attestation
procedure, and obtain an attestation response from the Measurement Engine. 
Specifically, \texttt{cfa\_init} initializes a new control-flow trace for attestation, and 
\texttt{cfa\_quote} generates the attestation response from initialized trace. 
Once \prv\ receives the challenge \emph{c} from \vrf, it initiates
a runtime trace by executing the \texttt{cfa\_init} function (\dTwo) and proceeds to execute the
target program. The attestation procedure on \prv\ may be initialized \emph{from within} the target
program in cases where an attestation of only a particular operation is desired. The target program
can, based on e.g., input from the verifier, decide \emph{at runtime} which portions of program
operation are to be attested. 


As a result of the instrumentation of the target program binary, all control-flow instructions are
intercepted by the Runtime Tracer Trampolines (\dThree). The Runtime Tracer determines the source
and destination address of the observed branch, and the type of control-flow instruction that caused
it (see Section~\ref{sec:instr}). It then triggers the Measurement Engine (\dFour). The Measurement
Engine performs the incremental measurement of the runtime control-flow path as described in
Section~\ref{sec:design}, i.e., computes the cumulative \auth\ (using the BLAKE2 hash function). The
Measurement Engine then transfers control back to the Runtime Tracer, which in turn resumes
the execution of the target program at the destination address of the observed branch (\dFive).
Once the program procedure that is to be attested completes, \prv\ executes the \texttt{cfa\_quote}
function (\dSix) with the challenge \emph{c} provided by \vrf. This triggers the Measurement
Engine to finalize \auth\ and generate the attestation response. 

\vspace{1mm}
\noindent\textbf{Trust Assumptions} are as follows:
\begin{compactitem}
  \item Assumption 1: The Normal World software stack, including the attested target program and Runtime
	Tracer components are covered by the static attestation from \prv\ to \vrf.
  \item Assumption 2: The Measurement Engine is trusted and cannot be disabled or modified. 
\end{compactitem}
Static attestation of the Normal World software stack ensures that \adv\ cannot modify this without
being detected by \vrf. This is important because it allows \vrf\ to detect if \adv\ tampers with
Normal World software to manipulate the inputs to the Hash Engine computing cumulative hashes.
Isolation of the Measurement Engine from the Normal World is important since, in the event of a
compromised control flow (even without modifying Normal World software), \adv\ might attempt to
influence hash computation prior to its delivery to \vrf.

It is also possible that an adversary may attempt to mount a runtime attack against the Runtime
Tracer and through this tamper the input to the Measurement Engine. However, the Measurement Engine
can prior to relinquishing control back to the Trampoline that triggered the measurement validate
that the actual destination of the recorded branch matches the input to the algorithm.    

We implemented the Measurement Engine as a monolithic trusted kernel executing in the ARM
TrustZone ``Secure World''~\cite{ARM-TrustZone}, thus isolating it from the untrusted target program,
and the Runtime Tracer, to prevent \adv\ from tampering with measurements. 

\noindent\textbf{Node IDs.}
Recall that \tool\ requires unique node IDs for CFG nodes. There are several options for doing this: 
{\bf (1): } Identify entry and exit addresses of basic blocks. These addresses are unique and already
present, without extra instrumentation. However, they change for each run of the program, if memory
layout randomization (ASLR)~\cite{sok-aslr} is used. {\bf (2):} Using labels, instrument the program to
embed a unique ID at the beginning of each basic block, similar to label-based CFI
schemes~\cite{Abadi2009}. {\bf (3):} Maintain a dedicated table that maps memory addresses of
basic blocks to node IDs. 

We selected {\bf (1)}, since many embedded systems do not yet
support ASLR due to lack of an MMU. However, if ASLR is used, we can either report the base address
in the attestation result, or adopt alternative approaches.

\subsection{ARM Instrumentation} \label{sec:instr}
\noindent
Another consideration for our implementation is the instrumentation of ARM binaries. 32-bit ARM processors feature 16 general-purpose registers. All these registers, including the program counter (\texttt{pc})
can be accessed directly. In addition to a 32-bit RISC instruction set, ARM processors also support a 16-bit
\emph{Thumb} instruction set. Thumb instructions act as compact shorthands for a subset of the 32-bit ARM instructions,
allowing for improved code density. Modern ARM processors extend the Thumb instruction set with 32-bit instructions
(distinct from those in the 32-bit ARM instruction set) that may be intermixed with 16-bit Thumb instruction. The
resulting variable-length instruction set is referred to as \emph{Thumb-2}.

Programs conforming to the ARM Architecture Procedure Call Standard (AAPCS)~\cite{ARM-AAPCS} perform subroutine calls
either through a \textbf{B}ranch with \textbf{L}ink (\texttt{bl}) or \textbf{B}ranch with \textbf{L}ink and
e\textbf{X}change (\texttt{blx}). These instructions load the address of the subroutine to the \texttt{pc} and the
return address to the link register (\texttt{lr}). The \emph{non-linking} variants, \textbf{B}ranch \texttt{b} and
\textbf{B}ranch and e\textbf{X}change, are used to direct the control flow within subroutines as they update the
\texttt{pc}, but leave \texttt{lr} untouched. The ARM architecture provides no dedicated return instruction.  Instead,
any instruction that is able to write to the program counter may function as an effective return instruction.

In order to faithfully reproduce the control flow of the program being attested, our instrumentation approach needs to
be able to intercept all instructions sequences that affect the control flow of the monitored program. On our evaluation
platform programs execute solely in the 32-bit ARM state, allowing us to instrument the program binary such that all
control-flow instructions are replaced by instructions that transfer control to a fixed piece of code belonging to the
Runtime Tracer. 

\noindent
\textbf{Binary rewriting.} In the instrumentation phase, the original branch targets for direct
branch instructions are collected from the target program binary to be stored at runtime in the read-only
branch table data structure that resides in target program memory.  Each target address is indexed
in the branch table by the location of the original branch instruction. These locations, the
locations of indirect branches, as well as other control-flow instructions are overwritten with
\emph{dispatch instructions} that redirect program control flow to a short piece of assembler code,
the \emph{trampoline}. We use the \texttt{bl} instruction as the dispatch instruction. The value of
\texttt{lr} after the dispatch is used to identify the call site of the trampoline. Our tool
utilizes the Capstone disassembly engine\footnote{\url{http://www.capstone-engine.org/}} to identify
control-flow instructions for rewriting.

\noindent
\textbf{Trampolines.} Each different type of control-flow instruction requires a separate trampoline, but unlike previous
work~\cite{MoCFI}, our approach does not leverage a separate trampoline for each instrumented branch. 
The trampoline manages the return address register as well as the register holding the destination address in
indirect branches. If the original instruction was a non-linking branch, the trampoline must
retain the original link register value from the prior \texttt{bl} originally present in the executable, and restore the
\ifnotabridged
program link register (to the value it had before the dispatch instruction) upon returning to program code.
\else
program link register upon returning to program code.
\fi

In order to avoid overwriting values stored in \texttt{lr}, the target program must not use \texttt{lr} as a
general purpose register. This can be enforced at compile time using appropriate compiler options\ifnotabridged\footnote{For instance
GCC provides  the \texttt{-ffixed-lr} option \url{https://gcc.gnu.org/onlinedocs/gcc-4.6.1/gcc/Global-Reg-Vars.html}}\fi.

During program execution, the trampoline will collect the return address from \texttt{lr} and determine 
the original branch target. For direct branches, the trampoline will look up the original destination address from the
branch table. For indirect branches, the trampoline will consult the corresponding register holding the destination
address, and for instructions utilized as function returns, the trampoline will either read the destination address from the stored
\texttt{lr} value, or read the return address from the program stack. It will then invoke the Measurement
Engine through a Secure World transition, passing as parameters the source and destination address. After trampoline
execution resumes, control is transferred back to the target program, to the original branch destination. 

\noindent
\textbf{Conditional branches.} All conditional branch instructions in the ARM instruction set have a linking \texttt{bl}
variant that updates \texttt{lr}. During instrumentation, we use the \texttt{bl} variant to redirect execution to the
corresponding trampoline, but do not change the condition under which the branch is executed. Conditional branches that
are not taken must at runtime be inferred from the branch information collected during the instrumentation phase, or by
introspecting the program code. In \tool, we make the branch table available to the Measurement Engine, and infer
non-taken conditional branches based on the start of the previous basic block known to the Measurement Engine, and the
source address of the branch that caused us to enter the Secure World.

\section{Evaluation} \label{sec:evaluation}
\noindent 
We evaluated our design and prototype implementation by applying \tool\ to a real embedded application.
This section describes our case study application, explains the attestation results, and discusses runtime performance.
The practical attacks and mitigations we demonstrated using this case study are presented in Section~\ref{sec:security}.

\subsection{Case Study of a Cyber-Physical System} \label{sec:case-study}
\noindent
A \emph{syringe pump} is an electromechanical system designed to dispense (or withdraw) precise quantities of fluid.
It is used in a variety of medical applications, especially those requiring small quantities of medication to be administered relatively frequently.
Many research fields, including chemical and biomedical research, also use syringe pumps.
Commercial syringe pumps are usually expensive, especially if they have been certified, but there has recently been significant interest in producing low-cost open-source alternatives~\cite{Wijnen2014}.

\noindent\textbf{Functionality.} 
A syringe pump typically consists of a fluid-filled syringe, a controllable linear actuator (e.g., using a stepper motor), and a control system.
The control system's task is relatively simple and would typically be implemented on an MCU -- it translates the desired input value (e.g., millilitres of fluid) into the control output for the actuator (e.g., setting a digital IO line high for a calculated duration).
However, given its intended usage, this system must provide a very high degree of assurance that it is operating correctly.
In both commercial and open-source designs, the control system may be a weak point since it runs embedded software and accepts external input (potentially from remote sources).
Undetectable runtime attacks affecting either the control flow or critical data variables could have serious consequences.
This is therefore a prime use case for C-FLAT. 

\noindent\textbf{Open Syringe Pump.}
For this case study, we used \emph{Open Syringe Pump}, an open-source, open-hardware syringe pump design.\footnote{\url{https://hackaday.io/project/1838-open-syringe-pump}}
It is controlled by an Arduino MCU running an embedded application written in Arduino Script.
Since \tool\ is not yet available for this type of MCU, we ported the application to a Raspberry Pi, which provides the required ARM TrustZone extensions.
Other open-source syringe pump designs already use the Raspberry Pi as a controller (e.g., \cite{Wijnen2014}), but run the control software as an application on top of a Linux OS.
To retain the embedded nature of the controller, we chose to port Open Syringe Pump as a \emph{bare-metal} implementation, which removes potential vulnerabilities introduced by an OS.
Our port required only minimal changes to the source code (less than 25\% of the overall application) in order to adapt it from Arduino Script to C.
Neither the functionality nor the control-flow graph (CFG) of the application was changed.
In terms of functionality, this application reads user input via the serial connection, and then performs one of the following types of actions: 
\begin{compactitem}
\item \texttt{set-quantity}: a numeric input changes a state variable representing the quantity of fluid to be used;
\item \texttt{move-syringe}: a trigger input causes the syringe pump to dispense or withdraw the set quantity of fluid;
\end{compactitem}
These types of actions naturally have different control flows, but even for the same action, the exact control flow depends on the state variable (i.e., the user-defined quantity of fluid).
Overall, the compiled application has a total of 13,000 instructions, and its static CFG contains 332 unique edges, of which 20 are loops.

\subsection{Attestation Results} \label{sec:attestation-results}
\noindent
We evaluated the functionality of our prototype implementation by instrumenting the syringe pump controller application and attesting the critical section of this application.
Specifically, we attested the sequence of actions that take place immediately after an external input is provided. 
This ensures that any action which exercises the critical functionality (e.g., causes the syringe to move) is covered by the attestation, whilst eliminating irrelevant time periods when the state of the system does not change.
We supplied a range of different inputs and recorded the resulting \auth\ values.

For the \texttt{set-quantity} path, the maximum length of \auth\ was for any input value was 1527~bytes.
This path contains 16 loop invocations, and 18 unique loop paths.
Similarly, for the \texttt{move-syringe} path, the maximum \auth\ length was 1179~bytes (12 loop invocations and 14 unique paths).

As expected, whenever the same control-flow path was taken (e.g., \texttt{set-quantity} or \texttt{move-syringe}), the same final hash value was obtained.
These hash values therefore enable the verifier to determine the precise control-flow path that was actually executed.
In this application, these hash values remained the same irrespective of the quantity of fluid dispensed/withdrawn by the syringe.
The reason for this is that the state variable is only used to determine how many times a critical loop should be executed.
The value of the state variable is therefore captured entirely in the additional loop information provided with the attestation.
Although the purpose of this loop handling behavior is to overcome the exponential increase in complexity caused by loops, in this application (and others like it) this also greatly simplifies the attestation output.

\subsection{Performance Impact} \label{sec:performance}
\noindent
We measured the time taken to perform \tool\ attestation in the syringe pump case study in order to determine (a) which factors affect the attestation overhead, and (b) whether this overhead is tolerable in real applications. 
All timing measurements were obtained using the hardware's built-in performance counters.
It is important to emphasize that, in a real-world deployment, this type of attestation would not be used on every execution of the program.
The verifier can choose when and how frequently to attest the prover.

\begin{figure}[ht]
\includegraphics[width=\linewidth]{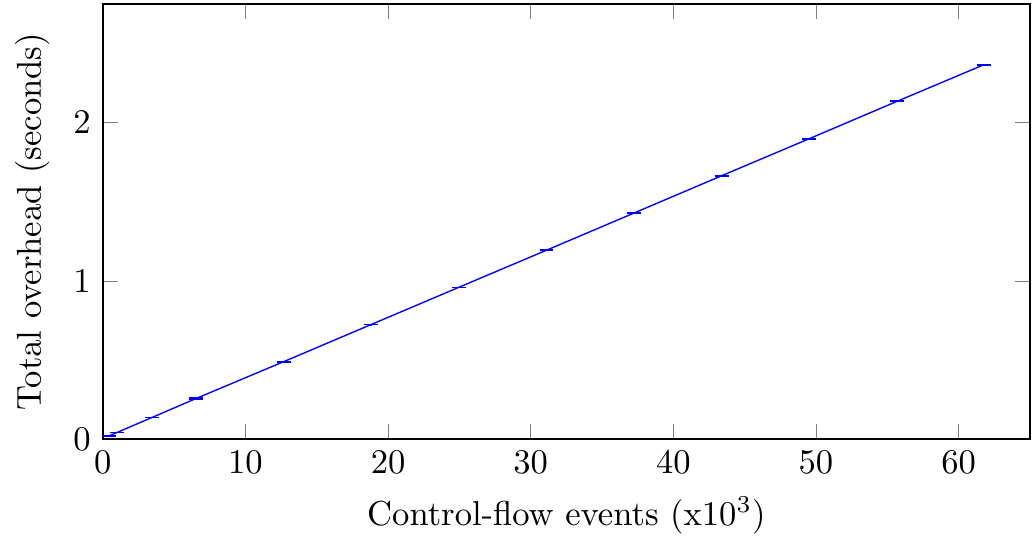}
\caption{Attestation overhead (average over 20 repetitions)}
\label{fig:overhead}
\end{figure}

\noindent
\textbf{Attestation overhead.}
We measured the time required to perform a complete attestation of the system, whilst varying the application's state variable through a representative range of values (i.e., changing the fluid quantity from 1~$\mu$L to 1000~$\mu$L).
We use the term \emph{control-flow event} to refer to any execution of a control-flow instruction (e.g., branch, return, etc.).
In other words, a control-flow event is a traversal of an edge in the CFG.
Since \tool\ executes its algorithm on every control-flow event, the overall attestation time varies depending on the number of control-flow events, as shown in Figure~\ref{fig:overhead}.

The attestation overhead is linearly proportional to the number of control-flow events.
This is as expected, since each control-flow event results in a trampoline invocation, a context switch, and an execution of the relevant Secure World algorithm.
The type of event also impacts the overhead (e.g., loop handling is more costly), but this is not visible in Figure~\ref{fig:overhead} due to the large number of events.
The variance of these measurements is negligible, since there is no other software running on the platform.

The attestation overhead can be decomposed into three categories: i) trampolines, ii) context switches, and iii) the Secure World algorithm.
For both the \texttt{set-quantity} and \texttt{move-syringe} paths, the relative percentage composition of the overhead remained roughly constant, irrespective of the state variable.
For the \texttt{set-quantity} path, 8.3\% of the overhead came from the trampolines, 11.9\% from the context switches, and 79.8\% from the algorithm itself.
These ratios were similar for the \texttt{move-syringe} path at 10.2\%, 17.5\%, and 72.3\% respectively.
As expected, the majority of the overhead is due to the algorithm itself. 

%
%

\noindent
\textbf{Performance impact.}
For the intended class of applications, this overhead would not necessarily have an impact on the real-world performance of the application.
In cyber-physical systems, the main timing requirements usually arise from the system's physical processes and components (e.g., sensors and actuators), which operate on comparatively slow physical timescales.
For example, in medical applications, typical flow rates for a syringe pump system could be $0.5\, ml/h$, $1.0\, ml/h$, and $2.0\, ml/h$.\footnote{\url{http://www.ncbi.nlm.nih.gov/pubmed/11032282}}
In our case study, the attestation overhead is $1.2\, s$ for dispensing $0.5\, ml$,  $2.4\, s$ for $1.0\, ml$, and $4.8\, s$ for $2.0\, ml$.
Therefore, the real-world attestation overhead for these parameters ranges from 0.03\% to 0.13\%.
Timescales in the order of minutes or hours are common in cyber-physical systems, for example, a typical embedded humidity sensor could measure and report a value once every 15~minutes\footnote{\url{http://vanderleevineyard.com/1/category/vinduino/1.html}}, and an embedded thermostat would not switch on a compressor unit more than once every three minutes.\footnote{\url{https://github.com/bbustin/climaduino/tree/develop}}

Even if the attestation overhead begins to affect the system's timing requirements, various strategies can be used to ameliorate this situation.
Firstly, we expect that the attestation would be run only occasionally, at a time determined by the verifier.
When the application is not being attested, the overhead decreases significantly (e.g., by 72\% to 80\% for the syringe pump).
This strategy could also be applied to other types of systems (e.g., attesting an embedded system in a car while not in use).

Secondly, the application could be adapted to accommodate this additional overhead.
The above measurements were obtained by applying C-FLAT to an unmodified application, but if the source code is available, minor modifications could be made to the application to compensate for the attestation overhead (e.g., reducing the durations of application-defined delay loops).

\noindent
\textbf{Implementation considerations.}
An unavoidable side-effect of \tool\ is that it may affect the way an application measures time.
If the application relies on time-calibrated loops to delay for a specific duration, the calibration will change when the application is instrumented.
Since the time spent in the Secure World varies depending on the type of control-flow operation, more complicated and/or experimental calibration would be required.
However, this can be easily avoided by instead using well-defined timing capabilities, such as system timers.

\subsection{Second Case Study} \label{sec:temperature}
\noindent
To further evaluate \tool, we completed a second case study using a significantly larger application.

We applied \tool\ to a soldering iron temperature controller, which allows users to specify a desired temperature and maintains the soldering iron at that temperature using a proportional-integral-derivative (PID) controller.\footnote{\url{https://create.arduino.cc/projecthub/sfrwmaker/soldering-iron-controller-for-hakko-907-8c5866}}
This type of system could also be used to control temperature in other settings, such as a building or an industrial process, and thus is representative of a large class of cyber-physical systems.
This application is significantly larger than the syringe pump as it consists of 70,000 instructions and has 1,020 static control-flow edges of which 40 are loops.

The security-critical functionality is the PID controller, which continuously reads the current temperature, compares it to the desired temperature, and calculates the PID output to adjust the temperature.
We instrumented the application to attest the control flow of this critical section, and verified that the correct hash values were obtained.
The total overhead added by \tool\ to a single iteration of the PID controller was $237\, {\mu}s$ (average over 100 runs) with a standard deviation of $1\, {\mu}s$.
As before, this overhead is significantly lower than the physical timescales on which these cyber-physical systems operate (e.g. attesting the controller once per second results in 0.03\% overhead).
For this attestation, the maximum length of \auth\ was 490~bytes.

\section{Security Considerations} \label{sec:security} \label{sec:eval-theoretical}
\noindent
In this section, we evaluate the security guarantees provided by \tool\ based on practical embedded exploits and different 
runtime exploitation techniques. Note that our discussion refers to the runtime attack threat model 
outlined in Figure~\ref{fig:runtime-attacks}.
The security of \tool\ itself is described alongside the implementation in Section~\ref{sec:implementation}.
As explained in that section, the integrity of the Runtime Tracer is ensured through static attestation, 
and the integrity of the measurement engine is protected by the trust anchor (e.g., the TrustZone Secure World).
As usual, the freshness of the attestation is guaranteed by the challenge-response protocol between 
\prv\ and \vrf\ (as described in Section~\ref{sec:model}).

\noindent\textbf{Exploiting the syringe program.} \label{sec:eval-practical}
We constructed three exploits that affect the \texttt{move-syringe} function.
Our first attack exploits a self-implanted memory error to trigger the movement of the syringe pump although 
no movement has been requested. As a consequence, liquid is dispensed at an unexpected time. We 
implemented this attack by reverse-engineering the syringe program binary and searching for return-oriented gadgets 
that allow us to load a function argument into ARM register \texttt{r0} and transfer control 
to the \texttt{move-syringe} function. We were able to construct a suitable gadget chain to implement 
this attack. 
With \tool\ enabled, we could 
detect this attack since the gadget chain leads to an unexpected hash measurement indicating to 
the verifier that a malicious program path has been executed. 

Our second exploit is based on a non-control-data attack: it corrupts a local state variable that controls 
the amount of liquid to use. This variable is used to determine the number of iterations of the critical loop 
in order to dispense the correct amount of fluid. 
Hence, this attack does not violate the program's CFG. 
However, for the case of the syringe program, the loop iterations indicate to the 
verifier how much liquid has been dispensed. Again, we can detect this attack with \tool\ since the precise loop 
iteration counts are reported in the attestation response. 


Our third exploit, a non-control-data attack, corrupts the static key-map array used for processing input from the keypad.
When a key is pressed, the keypad produces an analog value in a pre-defined range, which is read by the program.
The program iterates through a key-map array of known ranges, checking whether the analog input is the range for any recognized key.
By changing the pre-defined ranges in the key-map array, our attack misleads the program into believing that a physical key has been pressed, thus causing the program to perform the relevant action (e.g. the \texttt{right} key triggers the \texttt{move-syringe} function, causing the syringe to dispense liquid).
This attack can be detected by \tool\ because the number of iterations of the key processing loop reveals which key appears to have been pressed.
Assuming the verifier knows that the physical key has not been pressed, this can therefore be identified as unintended behavior.

\noindent\textbf{Code Injection.}
Conventional control-flow hijacking attacks rely on the injection 
and execution of malicious code~\cite{AlephOne96}. For this, the attacker needs to inject the malicious code into the 
data space of an application and corrupt control-flow information to redirect the execution to 
the injected code. Since malicious code resembles a new unknown node $N_X$ (see Figure~\ref{fig:runtime-attacks}), 
\tool\ can easily detect this attack. The reported hash value will include node $N_X$ 
although it does not belong to the original control-flow graph (CFG) of the application. To undermine 
\tool\ detection, an attacker may attempt to overwrite an existing node, e.g., replacing the 
benign Node $N_2$ with $N_X$. However, such an attack is not directly possible as we target 
platforms (see Section~\ref{sec:adv-model}) that enforce data execution prevention (DEP). Hence, the 
memory area of node $N_2$ is not executable and writable at the same time. Furthermore, the 
so-called multi-stage exploits which deploy code-reuse attack techniques in the first place to mark the code 
region as writable and inject malicious code thereafter are not feasible under \tool, because 
\tool\ will report the execution of the corresponding memory modification system calls (hereafter denoted as 
\emph{mem-calls}). 

The only possibility to launch 
a code injection attack requires a program that benignly \emph{mem-calls} to change memory 
permissions of code pages. For such cases, the attacker only corrupts input arguments to the \emph{mem-calls} so that 
node $N_2$ is eventually overwritten with $N_X$. These attacks resemble pure data-oriented 
exploits since they do not require any code pointer corruption. As such, they cannot directly be detected by means of 
control-flow attestation. On the other hand, if the injected code performs unexpected calls to benign code, \tool\ can still detect this pure data-oriented attack since an unknown hash value would be 
calculated.

\noindent\textbf{Return-oriented programming.}
Code-reuse attacks that only maliciously combine benign code snippets from different parts of 
the program are more challenging to detect because they do not inject any new nodes. 
The most prominent instantiation is return-oriented programming~\cite{Sh2007}. It combines short code sequences, 
each ending in a return instruction, to generate a new malicious program. \tool\ is able to detect 
return-oriented programming (including just-in-time return-oriented programming~\cite{JIT-ROP}) 
since unknown control-flow edges are taken upon execution of a return. 

In general, these attacks exploit return instructions to transfer control to either 
(i)~an arbitrary instruction in the middle of a CFG node, 
(ii)~the beginning of an arbitrary CFG node to which no valid control-flow edge exists, e.g., the 
attack path from $N_3$ to $N_2$ in Figure~\ref{fig:runtime-attacks}, or
(iii)~another valid CFG node that is not the actual caller, i.e., originating node. 
Case~(i) is detected by \tool\ since no valid measurement can be generated when a node 
is not executed from its original beginning; even for nodes to which valid control-flow edges 
exist. For instance, if an attacker exploits the vulnerability in $N_3$ to target an instruction in the 
middle of the otherwise valid node $N_6$, the reported hash value will be different to the one expected. 
This is due to the fact that our measurement includes source and end address of each node visited. 

For Case~(ii), the attacker follows a control-flow path that is not part of the CFG, e.g., node $N_3$ 
has no connection to node $N_2$. Due to the missing control-flow edge, the measurement will 
output an unknown hash value.

The last case is challenging to handle since the attacker targets a valid CFG node, i.e., there exists a valid 
control-flow edge but the current context indicates that the control-flow edge should not have been 
taken. We already discussed such scenarios in Section~\ref{sec:challenges} for the call-return matching. 
\tool\ indexes control-flow edges for function calls and returns to detect call-return mismatches. 
While this strategy works without special effort for direct function calls, indirect calls still pose an 
extra obstacle since static analysis may fail in resolving their target nodes. For those calls that 
static analysis cannot resolve, one can either leverage dynamic analysis or do an over-approximation 
by allowing each return instruction to target the instruction right after an unresolved indirect call. 
Note that this trade-off is not specific to our scheme, but a general 
problem in control-flow analysis. However, as we have shown in Section~\ref{sec:case-study}, embedded 
codes are typically tailored to a specific use-case allowing us to perform precise CFG analysis. In fact, 
basic dynamic analysis quickly allowed us to identify the valid paths in our syringe pump example.

\noindent\textbf{Jump-oriented programming.}
Another variant of return-oriented programming are jump-oriented programming 
attacks~\cite{rop-without-returns,JOP}. These attacks exploit 
indirect jump and call instructions at the end of each invoked code sequence. Similar to 
return-oriented programming, \tool\ detects these attacks since they deviate the program's 
control flow to an unintended CFG node. As long as static or dynamic analysis identifies the 
set of target addresses, we are able to determine valid measurements for each indirect jump and call 
thereby allowing \tool\ to detect malicious control-flow deviations incurred by jump-oriented programming.

\noindent\textbf{Function-reuse attacks.}
Function-reuse attacks invoke a malicious chain of subroutines. Typically, these attacks are leveraged to undermine 
mitigation schemes that specifically aim at detecting return-oriented programming attacks. 
\tool\ detects function-reuse attacks since it attests the program's entire control flow, i.e., it also captures the 
order in which functions are executing. For instance, the counterfeit object-oriented programming 
(COOP)~\cite{coop} exploitation technique abuses a loop gadget to invoke a chain of C++ virtual methods. \tool\ 
attests the execution within the loop, i.e., the number of times the loop has executed, and the order of virtual 
methods being called. Since the invoked chain does not resemble benign program execution, the 
resulting hash value will indicate to the verifier that a malicious path has been executed.

\noindent\textbf{Non-control-data attacks.}
Non-control-data attacks manipulate variables that affect the program's control flow~\cite{non-control-data}, e.g., to 
execute the privileged attack path (iii) in Figure~\ref{fig:runtime-attacks}. \tool\ is able to detect 
these attacks as each individual control-flow path leads 
to a different measurement. Hence, the verifier \vrf\ can immediately recognize which path has been taken 
by the program at the prover's device \prv. Another example is our attack reported in 
Section~\ref{sec:eval-practical} which only modifies a data variable to manipulate the amount of 
liquid dispensed through the syringe pump. Since the amount of liquid dispensed is proportional 
to the number of times a loop has executed, and that number is reported in the 
attestation response, \vrf\ can detect the malicious modification by validating the loop counters.

\section{Discussion} \label{sec:extensions}

\noindent
\textbf{Towards control-flow attestation of MCUs.} \label{sec:towards-mcus}
The
Raspberry Pi series of credit card-sized single-board computers have a become a popular
platform among developers, researchers and hobbyists for building embedded applications. Raspberry Pi~2 represents the higher-end of the embedded device spectrum. The
hardware trust anchor for the isolation of the \tool\ Measurement Engine is TrustZone-A, an architectural feature in ARM
application cores that has successfully been used as a hardware base for Trusted Execution Environments in mobile phones
for the last decade, and has also made inroads in the top-end Internet of Things (IoT) processor families. 

The Microcontroller (MCU) market for embedded computing in industry, automotive and home is rapidly moving towards
32-bit processing cores at the expense of small 8- and 16 bit controllers.
For the latest version of its MCU core line (ARM-M), ARM introduced TrustZone-M~\cite{ARMv8-M} -- hardware
primitives for isolating security-critical functionality such as firmware upgrades or secure channel establishment from
the rest of the controller software.
Intel's TrustLite~\cite{trustlite} provide very similar functionality. Both
TrustZone-M and TrustLite can provide the necessary hardware fundament to
implement \tool\ on IoT MCUs.

\noindent
\textbf{Reducing context switch overhead.} 
There are several optimization strategies depending on the properties of underlying secure hardware. On platforms
equipped with a full-featured TEE, the context switch into the Secure World is expensive. In the case of TrustZone-A on
an ARMv8 Cortex-A53 the context switch is approximately 3700 cycles or 5$\mu$s at 800MHz~\cite{TKVMPaol}. In the
TrustZone-A architecture, the transition between the Secure and Non-Secure modes occurs via hardware exceptions mediated
by a special exception handler referred to as the \emph{Secure Monitor}. The transition can only be triggered from the
Non-Secure World privileged mode. Under these circumstances, overall performance can be increased by caching control-flow
information in the Non-Secure World before transferring it in bulk to Measurement Engine in the Secure World.

Hardware security architectures for highly resource constrained embedded devices, such as TrustLite or TrustZone-M,
reduce the switching overhead associated with
trusted application invocation to a few cycles. For example, 
TrustZone-M has no need for a dedicated Secure Monitor context. The transition between Secure and Non-Secure Worlds is
handled by the processor circuitry independently of software through a mechanism based on protection regions with marked
call gates enforced by the Memory Protection Unit.

\noindent
\textbf{Embedded operating systems.} 
While the prover in our current prototype runs unassisted on bare-metal, many IoT devices feature
embedded operating systems. The main purpose of an embedded OS (e.g.,
FreeRTOS\footnote{\url{http://www.freertos.org/}},
Zephyr\footnote{\url{https://www.zephyrproject.org/}},
ChibiOS\footnote{\url{http://www.chibios.org}}) is to schedule tasks or functions, often based
mainly on interrupt input. The OS may also provide memory management for the tasks and support
debugging. A real-time OS adds pre-emptive scheduling to provide timing guarantees for individual
tasks. However, hardware in controllers traditionally do not support many processing contexts or
memory protection to achieve isolation between tasks or even between OS and application. From a
security perspective, \tool\ on such devices should therefore consider a single isolation
domain even if an OS is present, and instrument the OS as a single bare-metal program.

\noindent
\textbf{Thumb-2 instrumentation.} 
Modern ARM-M MCUs utilize the Thumb-2 instruction set because of its improved code density in order to maximize the usage
off on-chip Flash memory. In ongoing work, we enhance our instrumentation approach to include support for the
variable-length Thumb-2 instructions.

\noindent
\textbf{Data-flow attestation.} 
As already mentioned in Section~\ref{sec:bg-attacks}, we focus on control-flow related attacks. On the other 
hand, pure data-oriented attacks can lead to leakage of sensitive information, e.g., a data pointer that is 
processed in a network operation is altered to point to sensitive information. Hence, in our future 
work we will explore data-flow attestation, i.e., mechanisms that allow us to track data pointers and 
validate function arguments and return values.

\noindent
\textbf{Probe-based runtime attestation.} 
In our ongoing work, we explore control-flow attestation for larger and very complex programs, 
i.e., programs with a complex control-flow graph that contains a large number of different program paths. 
These might lead to a combinatorial explosion of valid hash values 
thereby rendering validation cumbersome. To support such programs, we envision probe-based 
attestation that (i)~generates and reports a cumulative hash for program segments, e.g., reporting 
an authenticator for each subroutine, and (ii)~allows probing a program during its execution. The 
latter enables an interactive attestation protocol in which \vrf\ can repeatedly validate the evolution 
of the program execution at arbitrary time intervals.

\section{Related Work} \label{sec:related}
\noindent
Prior work on static attestation falls into three main categories: (1) TPM-based, 
(2) software-based, and (3) hybrid. Although TPMs and subsequent techniques (e.g., \cite{flicker,McCune2010})
are pervasive in PCs, they are not well-suited for low-end embedded devices due 
to their relatively high cost. Meanwhile, software-based methods (e.g., SWATT~\cite{swatt}, 
Pioneer~\cite{Seshadri2005}, and VIPER~\cite{Li2011}) can be used -- under some strict assumptions -- 
for static attestation of certain legacy devices, such as peripherals.
Between these two extremes, hybrid schemes such as SMART~\cite{smart}, 
TrustLite~\cite{trustlite}, and TyTAN~\cite{Brasser2015} require minimal hardware trust anchors in order to 
provide strong \emph{remote} attestation guarantees. \tool\ is complementary to all these schemes in that it 
attests runtime behavior, which is orthogonal to static attestation.

Mitigation of runtime exploits is an ongoing research field with many solutions proposed 
over the last few years~\cite{eternal-war}. The main directions are control-flow integrity (CFI)~\cite{Abadi2009} 
and fine-grained code randomization~\cite{Co1993,sok-aslr}. The former ensures that a 
program only follows a valid path in the program's CFG. Many improvements have been made 
in the last years making CFI an efficient and effective mitigation technology~\cite{cfi-cots,CFI-forward:Ulfar}. 
On the other hand, CFI does not cover non-control-data attacks which lead to execution of unauthorized but 
valid CFG paths. The latter randomizes the code layout by randomizing the order of memory pages, 
functions, basic blocks, or instructions~\cite{sok-aslr}. However, the attacker can still exploit  
branch instructions to jump at her target address of choice. As such, code randomization cannot 
provide control-flow information which would allow \vrf\ to attest the program's execution. Another 
scheme is code-pointer integrity (CPI)~\cite{CPI} which aims at ensuring the integrity of 
code pointers. While it provides high assurance against control-flow hijacking attacks, 
it does not provide any information on the actual control-flow path taken. As such, similar to CFI, CPI does not 
cover non-control data attacks.

Property-based attestation explores behavioral characteristics beyond hash values of applications' 
binaries~\cite{propatt}. However, such schemes typically require a trusted third-party. 
Software-based attestation enables remote attestation for embedded devices 
without requiring a trust anchor~\cite{swatt}. The measurement is based on the calculation of a 
hash over the application's code memory, where the attestation is only successful if the prover 
responds in a certain time interval. 
However, this assumes no noise on the channel, and requires the hash function to be optimal. 
Semantic remote attestation enables attestation of program behavior by enforcing local policies at 
the Java bytecode layer~\cite{Franz2004}. 
Unlike \tool, neither property-based, semantic, nor software-based attestation cover control-flow attacks 
at the binary level.

There also exist several approaches to enable dynamic remote attestation: ReDAS~\cite{redas} 
explores runtime properties such as the integrity of a function's base pointer. While checking 
against such properties enhances static remote attestation, it only captures specific application 
fields and pointers rather than the whole control-flow path. Trusted virtual 
containers~\cite{tvc-smith} allow control-flow attestation but at a coarse-grained level. That is, 
they only attest the launch order of applications modules, but not the internal control flows of 
an application. Lastly, DynIMA~\cite{DaviROP2009} 
explores taint analysis to achieve runtime attestation. However, its implementation only validates 
invariants on return instructions such as the number of instructions executed between two 
consecutive returns. 

\section{Conclusion} \label{sec:conclusion}
\noindent
Memory corruption attacks are prevalent on diverse computing platforms and can lead 
to significant damage on embedded systems that are increasingly deployed in safety-critical 
infrastructures. In particular, there is not yet a mechanism that allows precise verification of 
an application's control flow to detect such attacks. \tool\ tackles this problem by means of 
control-flow attestation allowing a verifier to detect control-flow deviations 
launched via code injection, code-reuse, or non-control-data attacks. Our prototype implementation 
on an ARM-based Raspberry Pi demonstrates that \tool\ can be leveraged to detect different runtime 
exploitation techniques launched against embedded software such as a syringe pump program, i.e., \tool\ detects when 
an attacker manipulates the amount of liquid dispensed through the syringe. 
The source code of \tool\ is available at \materials.

\section{Acknowledgments} \label{sec:acknowledgements}
\noindent
This work was supported in part by the Intel Collaborative Institute for Secure
Computing at TU Darmstadt and Aalto University. This work
was also supported in part by the Academy of Finland (283135), Tekes (1226/31/2014), the German Science Foundation (project
S2, CRC 1119 CROSSING), the European Union's Seventh Framework Programme (643964, SUPERCLOUD), and the
German Federal Ministry of Education and Research within CRISP. At UC Irvine, this research was
supported by funding from the National Security Agency (H98230-15-1-0276) and the Department of
Homeland Security (under subcontract from the HRL Laboratories).

\bibliographystyle{abbrv}
\bibliography{cfi} 

\end{document}